
\magnification=1200
\vsize=8.5truein
\hsize=6truein
\baselineskip=14truept
\parindent=0.5truecm
\font\scap=cmcsc10
\font\pieni=cmr8

\hfuzz=0.8truecm
\def\hdot{\!\cdot\!}
\def\frac#1#2{\textstyle{#1 \over #2}}

\null \bigskip
\centerline{\bf INTEGRABLE TRILINEAR PDE's}\footnote{}{To appear in the
proceedings of NEEDS'94 (11-18 September, Los Alamos)}

\vskip 0.8truecm
\centerline{\scap J. Hietarinta}
\centerline{\sl Department of Physics, University of Turku,
20500 Turku, Finland}
\vskip 0.8truecm
\centerline{\scap B. Grammaticos}
\centerline{\sl LPN, Universit\'e Paris VII,
Tour 24-14, 5${}^{\grave eme}$ \'etage, 75251 Paris, France}
\vskip 0.8truecm
\centerline{\scap A. Ramani}
\centerline{\sl CPT, Ecole Polytechnique, 91128 Palaiseau, France}
\vskip 0.8truecm

\centerline{\scap Abstract}
{\baselineskip=12truept
\narrower\smallskip\noindent{\pieni
In a recent publication we proposed an extension of Hirota's bilinear
formalism to arbitrary multilinearities. The trilinear (and higher)
operators were constructed from the requirement of gauge invariance
for the nonlinear equation.  Here we concentrate on the trilinear
case, and use singularity analysis in order to single out equations
that are likely to be integrable. New PDE's are thus obtained, along
with others already well-known for their integrability and for which
we obtain here the trilinear expression.}\smallskip }

\baselineskip=14truept
\vskip 0.6truecm
\noindent{\bf 1. Introduction}
\medskip
\noindent What is astonishing with integrable PDE's is that there are so
many of them.  Infinite hierarchies of equations are known to date and
the domain is still expanding.  Moreover, these equations are often
associated to physical models [1].  Then, using integrability, one can
construct particular solutions and compute the pertinent physical
quantities for these models.

One of the essential tools in the study of integrable PDE's, over the
past 20 years, has been the bilinear formalism of Hirota [2]. The main
advantage of this approach lies in the fact that it allows one to
obtain multisoliton solutions in a straightforward way [3]. Still, the
proof of integrability, based on the existence of an arbitrary number
of solitons, may present considerable difficulties. It is often easier
to test the integrability of a given equation using singularity
analysis (Painlev\'e method)[4], especially since this criterion
(absence of singular solutions that exhibit branching) can be
implemented algorithmically.

In a recent work [5] we have presented an extension of Hirota's
bilinear formalism to higher multilinearities. This was motivated by
the existence of integrable equations that cannot be bilinearized,
like the Satsuma determinantal trilinear PDE's [6]. (It must be made
clear at this point that the nonexistence of a bilinear form is not a
rigorous statement; it simply means that no bilinear form has been
obtained to date). Our method is based on the assumption of gauge
invariance. In the first exploratory study [5] we studied only the
leading part of one-dimensional trilinear equations with one dependent
variable and found those that satisfy the Painlev\'e criterion. The
aim of the present paper is to complete this work, and obtain the
integrable higher dimensional generalizations with the non-leading
parts.  As we will show in the following sections some of the
equations obtained are new, while others are well-known for their
integrability without necessarily possessing a simple bilinear form.

\vskip 0.6truecm
\noindent{\bf 2. From bilinear to trilinear operators.}
\medskip
\noindent
A prerequisite to the application of the Hirota bilinear formalism is
a dependent variable transformation that converts the nonlinear
equation into a quadratic `prepotential' form. In order to make things
more clear let us present the classical example of the KdV equation.
Starting from
$$u_{xxx}+6uu_x+u_t=0,\eqno(2.1)$$
we introduce the transformation $u=2\partial^2_x\log F$ and obtain
(after one integration):
$$F_{xxxx}F-4F_{xxx}F_{x}+3F^2_{xx}+F_{xt}F-F_xF_t=0.\eqno(2.2)$$
This last equation can be written in a particularly condensed form
using the Hirota $D$ operator:
$$(D^4_x+D_xD_t)F\hdot F=0,\eqno(2.3)$$
where the $D$-operator is defined by its antisymmetric derivative
action on a pair of functions (the `dot product'):
$$D_x^n f\hdot g=(\partial_{x_1}-\partial_{x_2})^nf(x_1)g(x_2)
\big |_{x_2=x_1=x}.\eqno(2.4)$$

The crucial observation here is the relation of the `physical'
variable $u=2\partial^2_x\log F$ to the Hirota's function $F$: the
gauge transformation $F\to e^{px+\omega t} F$ leaves $u$ invariant. It
turns out that this is a general property of bilinear equations. In
fact, one can {\sl define} Hirota's bilinear equations through the
requirement of gauge invariance. This statement was proven in [5].

Having obtained the Hirota bilinear operators on the sole requirement
of gauge invariance we investigated in [5] the possible extension of
this formalism and the introduction of multilinear operators. This
turned out to be possible.  Our first step was the extension of the
bilinear operators to the trilinear case.  We found that one
convenient basis for the representation of the gauge invariant $N$'th
order derivative operators is given by
$(\partial_1-\partial_2)^n(\partial_1-\partial_3)^{N-n}$ for
$n=0,\dots,N$, where the subscripts tell on which of the three
functions each derivative acts.  Thus the basic building blocks for the
trilinear operators are again the Hirota bilinear $D$'s: we must just
specify the indices in this case. We thus have
$D_{12}\equiv\partial_{x_1} -\partial_{x_2}$,
$D_{23}\equiv\partial_{x_2} -\partial_{x_3}$,
$D_{31}\equiv\partial_{x_3} -\partial_{x_1}$, but, of course the three
are not linearly independent: $D_{12}+D_{23}+D_{31}=0$. Their action
on a `triple dot product' is analogous to the bilinear case:
$$D_{12}f\hdot g\hdot
h=(\partial_{x_1}-\partial_{x_2})f(x_1)g(x_2)h(x_3)\big
|_{x_3=x_2=x_1=x}=(f'g-fg')h.\eqno(2.5)$$

The choice of a particular pair of $D$'s as the basic trilinear
operators breaks the symmetry between the three coordinates
$x_i$'s. It is possible to restore this symmetry by introducing a
different basis for the trilinear operators, $T$ and $T^*$:
$$T=\partial_1+j\partial_2+j^2\partial_3\, , \quad
T^*=\partial_1+j^2\partial_2+j\partial_3,\eqno(2.6)$$ where $j$ is the
cubic root of unity, $j=e^{2i\pi/3}$. (Note that the star in $T^*$
indicates complex conjugation for the coefficients in $T$ but not for
the independent variables). Note that $T^nT^{*m}F\hdot F\hdot F=0$,
unless $n-m\equiv 0$ (mod 3), which is the equivalent to the bilinear
property $D^n F\hdot F=0$, unless $n\equiv 0$ (mod 2). Moreover, a
reality condition is satisfied: $T^nT^{*m}F\hdot F\hdot
F=T^mT^{*n}F\hdot F\hdot F$.

The generalization to higher multilinear equations is
straightforward. One can introduce the set of $n(n-1)/2$ operators
$D_{ij}$ acting on $n$-tuple dot-products $D_{ij}f_1\hdot f_2\!\cdot
\dots \cdot\! f_n.$ Of course only $n-1$ of the $D_{ij}$'s are
independent, a convenient basis being the $D_{1j},$ $j=2, \dots n$. As
in the trilinear case, one can also construct `symmetric' operators:
$$M_n^m=\sum_{k=0}^{n-1} e^{2\pi ikm/n} \partial_{k+1}\eqno(2.7)$$ for
$0<m<n$. We have, for example, $D=M_2^1$, $T=M_3^1$, $T^*=M_3^2$ and
so on.

\vskip 0.6truecm
\noindent{\bf 3. General comments on multilinearization}
\medskip
\noindent
Before dealing with specific cases let us present here some general
considerations.  Let us start with a nonlinear (in $u$) equation, of
order $k+1$ having the form $u_t+\partial_x
P(u,u_x,\dots,u_{kx})=0$. Several well known integrable equations
belong to this class.  Let us assume that the leading part of $P$ is
weight-homogeneous in $u$ and $\partial_x$, with $u$ having the same
weight as $\partial_x^2$. Then we can transform the equation into a
multilinear expression through the transformation $u=\alpha
\partial_x^2(\log F)$ and obtain, after one $x$-integration,
generically an $(k+2)$-multilinear equation. The scaling factor
$\alpha$ can then be chosen (perhaps in several ways) so as to make an
$F^2$ term factor out: the resulting multilinear equation is at most
$k$-linear.  For example at order five ($k=4$) we have three
integrable equations, and we should expect, in principle, these
equations to have quadrilinear forms. Some unexpected cancellations,
however, do occur. Thus the Sawada-Kotera equation [7] has a bilinear
expression, the Lax-5 equation [8] a trilinear form, but the
Kaup-Kuperschmidt equation [9] is quadrilinear.

In the process of multilinearization of a given nonlinear equation the
following points should be noted:

1) Above we discussed only $u=\alpha\partial^2 \log(F)$ substitutions. If
$u$ always appears in the equation with at least one derivative then we
can get a gauge invariant form also with $u=\alpha\partial \log(F)$,
and if we always have at least two derivatives, $u=\alpha \log(F)$ is
sufficient.

2) Any bilinear expression multiplied by $F$ is cubic and gauge
invariant and therefore has trilinear form, for example
$$\displaylines{
\qquad F(D_xD_y )F\hdot F=\frac{2}{3}\, T_xT_y^* F\hdot F\hdot F,
\hfill(3.1)\cr
\qquad F(D_x^3D_y) F\hdot F=\frac{2}{3}\, T_x^2T_x^*T_y^*
F\hdot F\hdot F,\hfill(3.2)\cr
\qquad F(D_x^5D_y )F\hdot F=\frac{2}{27}\,(10T_x^3T_x^{*2}T_y^*- T_x^5T_y)
F\hdot F\hdot F,\hfill(3.3)\cr
\qquad F(D_x^7D_y )F\hdot F=\frac{2}{81}\,(35T_x^4T_x^{*3}T_y^*
-T_x^7T_y^*-7T_x^6T_x^*T_y) F\hdot F\hdot F.\hfill(3.4)}$$
 Therefore known one-component bilinear
equations reappear in this trilinear study.

3) Taking a derivative of an equation having an $n$-linear form yields an
equation that is $n+1$ multilinear, for example
$$\displaylines{
\qquad\partial_x\left(F^{-2}D_xD_y F\hdot F\right)=
\frac{2}{3}F^{-3}\,T_x^2T_y F\hdot F\hdot F,\hfill(3.5)\cr
\qquad\partial_x\left(F^{-2}D_x^3D_y F\hdot F\right)=\frac{2}{9}F^{-3}\,
\left(T_x^4T_y^*+2T_x^3T_x^*T_y\right) F\hdot F\hdot F,\hfill(3.6)\cr
\qquad\partial_x\left(F^{-2}D_x^6 F\hdot F\right)\phantom{D_y}=
\frac{2}{3}F^{-3}\, T_x^5T_x^{*2} F\hdot F\hdot F.\hfill(3.7)}$$
The integrability of the
derivative form is another matter: if for example the $x$-derivate form
is also integrable then the original equation would be integrable with
inhomogeneous $x$-independent term. Thus we will later find the
$x$-derivative of Ito equation to be integrable, but not its
$t$-derivative, therefore Ito's equation is integrable even with a
inhomogeneous $g(t)$-term.

Since equations having a bilinear form will reappear in the trilinear
analysis as mentioned before, let us collect here the integrable
equations that can be bilinearized to the form $P(D)F\hdot F=0$ [10,11]:

Kadomtsev-Petviashvili:
$$ \left(D_x^4-4D_xD_t+3D_y^2\right)F\hdot F=0,\eqno(3.8)$$

Ito:
$$ \left(D_tD_x^3+aD_x^2+D_tD_y\right)F\hdot F=0,\eqno(3.9)$$

Hietarinta:
$$ \left(D_x(D_x^3-D_t^3)+aD_x^2+bD_tD_x+cD_t^2\right)F\hdot F=0,
\eqno(3.10)$$

Sawada-Kotera-Ramani:
$$ \left(D_x^6+5D_x^3D_y-5D_y^2+D_xD_t\right)F\hdot F=0,\eqno(3.11)$$

\vskip 0.6truecm
\noindent{\bf 4. Singularity analysis of trilinear equations}
\medskip
\noindent
One motivation behind the multilinear approach is to find new
integrable equations. Eventually we would like to get something
similar to the systematic classification of bilinear
equations presented in [10,11]. In the following
paragraphs we will limit ourselves to the singularity analysis of
trilinear equations involving only one {\sl dependent} variable, i.e.
unicomponent equations $P(T_x,T_x^*,T_y,T_y^*,\dots)F\hdot
F\hdot F=0$. (Let us recall here that in the bilinear case a complete
classification of these simplest equations [10,11] was possible).

Since the dependent function $u$ in a nonlinear equation is related to
the multilinear $F$ through $u=\alpha\partial^2\log F$ it is clear
that a zero in $F$ induces a pole-like behavior in $u$.  Let us here
consider a concrete example: $T^{3k}F\hdot F\hdot F=0$.  Putting, for the
dominant part, $F\sim \phi^n$, where $\phi$ is the singular manifold
we find that the only possible leading behaviors have
$n=0,1,\dots,k-1$. The first corresponds to a nonsingular Taylor-like
expansion, which is always possible. The second behavior $F\sim\phi$
corresponds to a simple zero of $F$ that would give a (double) pole in
$u$.  Next the resonances $r$ can be obtained if we substitute
$F=\phi^n(1+\omega\phi^r)$ and collect terms linear in $\omega$. For
the equation to pass the Painlev\'e test the resonances must be
integers and no incompatibilities must arise at any resonance.  As an
example let us take $k=2$, i.e., the operator $T^6$. The
leading behavior $F\sim\phi$ leads to the resonances $r=-1,0,1,2$ and
the roots of $r^2-13r+60=0$ which are complex. Thus $T^{6}F\hdot F\hdot
F=0$ does not possess the Painlev\'e property.

\vskip 0.4truecm
\noindent {\sl 4.1 Leading behaviors}

\noindent The Painlev\'e analysis for the leading part of the
equations was performed in [5], with just one independent variable.
This is sufficient for the computation of dominant singularities and
resonances.  Note that for a given $N=n+m$ there may exist several
pairs $(n,m)$ such that $T^nT^{*m}F\hdot F\hdot F$ is not identically
zero, namely those for which $n\equiv m$ (mod 3). The leading part of
the general equation at order $N$ is then given by a linear
combination of all the non-vanishing $T^nT^{*m}F\hdot F\hdot F$'s. In
each case we give below the precise combinations that lead to
equations with the Painlev\'e property. The nonlinear forms of the
leading parts of the equations are obtained by the standard
substitution $F=e^g$ followed by $u=2g''$. The results of [5] are
summarized as follows:

\noindent ${\bf N=2}: \quad{\rm operator:}\ TT^*\quad
{\rm leading\ part:}\ u$.
\item{} In this case we can write the result also in bilinear form

\noindent ${\bf N=3}: \quad{\rm operator:}\ T^3\quad
{\rm leading\ part:}\ u'$.

\noindent ${\bf N=4}: \quad {\rm operator:}\ T^2T^{*2}\quad
{\rm leading\ part:}\ u''+3u^2$.
\item{}This, of course, is just the leading part of the KdV equation
in potential form, and can also be written in bilinear form, see (3.2).

\noindent ${\bf N=5}: \quad {\rm operator:}\ T^4T^*\quad {\rm
leading\ part:}\ u'''+6u'u$.
\item{} This is the derivative of the expression obtained at $N=4$.

\noindent ${\bf N=6}: \quad {\rm operator:}\ \lambda T^6 +\mu T^3T^{*3}$.
This is the first case where we have two possible $(n,m)$ pairs. The
$\lambda,\mu$ combinations that pass the Painlev\'e test are the
following
\item{a)} $(\lambda=7, \mu=20)\quad {\rm leading\ part:}\
u''''+10u''u+5u'^2+10u^3$.
\item{} This is the leading part of the  once integrated 5th order
equation in the Lax hierarchy of KdV.
\item{b)} $(\lambda=-2, \mu=20)\quad {\rm leading\ part:}\
u''''+15u''u+15u^3$.
\item{} This is the leading part of the  once integrated Sawada-Kotera-Ramani
equation (3.11), and can also be written in bilinear form, see (3.3).

\item{c)} $(\lambda=-\mu)\quad {\rm leading\ part:}\ uu''-u'^2+u^3$.
\item{} This expression can be cast in determinantal equation
$$\left|\matrix{ F''''&F'''&F''\cr F'''&F''&F'\cr F''&F'&F\cr }\right|=0,$$
whose integrability was already noticed by Chazy [12].

\noindent ${\bf N=7}: \quad {\rm operator:}\ T^5T^{*2}\quad
{\rm leading\ part:}\  u^{(5)}+15u'''u+15u''u'+45u'u^2$.
\item{} This is the Sawada-Kotera equation, i.e., the derivative of the
expression obtained in $N=6$b, c.f. (3.7).

\noindent ${\bf N=8}: \quad {\rm operator:}\ \lambda T^7T^* +\mu
T^4T^{*4}$ and
nonlinearization with  $u=6g''$ instead of $u=2g''$ used before.
\item{a)} $(\lambda=4, \mu=5)\quad {\rm leading\ part:}\
u^{(6)}+6u''''u+10u'''u'+5u''^2+10u''u^2+10u'^2u+{5\over 3}u^4$
\item{} This would correspond to a $7^{th}$ order equation which, we
believe, leads to a new
integrable case.
\item{b)} $(\lambda=4, \mu=14)\quad {\rm leading\ part:}\
u^{(6)}+7u''''u+7u'''u'+7u''^2+14u''u^2+7u'^2u+{7\over 3}u^4$
\item{} This is the leading part of a once integrated higher Sawada-Kotera
equation.
\item{c)} $(\lambda=-\mu)\quad {\rm leading\ part:}\
u''''u-3u'''u'+2u''^2+4u''u^2-3u'^2u+{2\over3}u^4$.
\item{} This new equation looks like an extension of Satsuma's equation
given at $N$=6c above, but it cannot be written as a single determinant.

\noindent ${\bf N=9}$: \quad No cases passing the Painlev\'e test exist at
this order.

\noindent ${\bf N=10}: \quad {\rm operator:}\ 5 T^8T^{*2} +4 T^5T^{*5}$
(and we take $u=30g''$).
\item{} ${\rm leading\ part:}\
u^{(8)}+2u^{(6)}u+4u^{(5)}u'+6u''''u''+5u'''^2+{6\over
5}u''''u^2+4u'''u'u+2u''^2u+2u''u'^2
+{4\over 15}u''u^3+{2\over 5}u''^2u^2+{2\over 375}u^5$
\item{} This is also a new equation.

\noindent Furthermore, no integrable candidates were found at orders
$N=11,12$. The singularity analysis at these higher orders becomes
progressively more difficult.( Already at $N=12$ there exist three
non-vanishing $n,m$ combinations). It is, thus, not possible to extend
our investigation to very high orders (as was done in our study of
bilinear equations [11]), but we do believe that no further integrable
candidates exist at higher orders.

\vskip 0.4truecm
\noindent {\sl 4.2 Resonance conditions}

\noindent In order to investigate the resonance conditions we must specify
precisely the PDE we are working with, i.e., also the nonleading
parts. Let us state from the outset that we are not going to consider
equations of the form $N=6c$ or $N= 8c$. These equations have
a form that makes the {\sl expected} leading singularity to vanish
identically, which leads to certain complications in the singularity
analysis: they will be discussed elsewhere. Also in every case examined,
subleading parts of opposite parity (= odd vs.\ even number of $T$'s)
to that of the leading were found to violate the Painlev\'e property,
but we did not make an exhaustive study of this.  The technical
details of the resonance conditions study are not particularly
interesting, the analysis was performed on a computer using the REDUCE
language [13] for symbolic manipulation. Although in some cases the
resonances are particularly high (in the case of $N=10$ we have a
resonance at $r=30$ for the leading behavior $F\sim\phi^{14}$) we
were able in all cases to check the resonance conditions.

For low $N$ the complete results are simple: for $N=3$ one obtains only
linear equations, and for $N=4$ all the integrable equations could also
be written in bilinear form.

For equations with $N\ge 5$ an important question was how to extend
the leading-order trilinear operator to two dimensions and still keep
the calculations manageable. Our choice has been to consider only
`monomial' leading parts, that is monomial in derivatives and not
necessarily in $T$'s. (From our experience on the bilinear case we do
not expect nonmonomial leading parts to play a role for $N\ge 5$).

\noindent ${\bf N=5:}$

\noindent The leading 1-dimensional operator is $T_x^4T_x^*$ and
three monomial generalizations to 1+1 dimensions are possible: i)
$T_x^4T_x^*$, ii) $T_x^4T_y^*+\lambda T_x^3T_yT_x^*$ and iii)
$T_x^3T_yT_y^*+\lambda T_x^2T_y^2T_x^*$.

The case i) supplemented by subleading terms leads only to the
$x$-derivative of the KP equation (3.5-8):
$$(T_x^4T_x^*-4T_x^2T_t+3 T_xT_y^2)F\hdot F\hdot F=0.$$

In the case ii) (discarding the $\lambda=-1$ case that does not have a
standard leading singularity) there exist the three Painlev\'e
candidates $\lambda=-4,\, 2$ and 8. The first case does not satisfy
the resonance condition. The case $\lambda=2$ yields just the
$x$-derivative of the Ito equation (3.9):
$$(T_x^4T_t^*+2T_x^3T_x^*T_t+3aT_x^3+3T_x^2T_y)F\hdot F\hdot F=0.$$
The final case $\lambda=8$ is more
interesting. The most general result passing the Painlev\'e test
reads:
$$\left(T_y(T_x^{*4}+8T_x^3T_x^*+9T_y^2)+9T_x^2T_t\right)F\hdot F\hdot
F=0.\eqno(4.1)$$
Putting $F=e^{g/2}$ we obtain:
$$g_{xxxxy}+4g_{xxy}g_{xx}+2g_{xy}g_{xxx}+g_{yyy}+g_{xxt}=0.\eqno(4.2)$$
This equation is a generalization (the $g_{yyy}$ term is new) of an
equation obtained by Schiff [14] from a reduction of the self-dual
Yang-Mills equations. Incidentally, for this equation we have checked
that a nontrivial three soliton solution does exist and this is a
further indication of its integrability. It is interesting to observe
that this equation reduces to the derivative KP of Case i) if the
first factor $T_y$ is replaced by $T_x$ in (4.1).

Finally the case iii) did not lead to any equations with the
Painlev\'e property.

A general study of monomial leading parts was performed for all $N$'s
higher than 5. To make a long story short, only a monomial leading
part of the form $T_x^{N-1}T_y$ has acceptable resonances for
$N=6,7$. However, even in these cases, the resonance conditions are
not satisfied.  Thus, beyond $N=5$ the leading part is not only
monomial but also 1-dimensional.

\noindent ${\bf N=6:}$

\noindent The leading
operator is $\lambda T_x^6+20T_x^3T_x^*$ with $\lambda =$7 or $-2$. For
$\lambda =7$ the equation obtained is the fifth-order equation in
the Lax hierarchy [8] plus the KdV: $$(20T_x^3T_x^{*3}+7T_x^6+\alpha
T_x^2T_x^{*2}+27T_x^*T_y)F\hdot F\hdot F=0\eqno(4.3)$$ or in nonlinear
form after one integration ($u=2\partial_x^2 \log F$):
$$u_{5x}+10uu_{xxx}+20u_xu_{xx}+30u^2u_x+\alpha
(u_{xxx}+6uu_x)+u_y=0\eqno(4.4)$$ Thus the Lax-5 equation does not
possess a simple bilinear form like KdV but a trilinear one.  For
$\lambda =-2$ the resulting equation is the 3-dimensional
Sawada-Kotera-Ramani equation,
$$(10T_x^3T_x^{*3}-T_x^6+45\alpha(T_x^2T_x^*T_y^*-\alpha
T_yT_y^*)+9T_x^*T_z)F\hdot F\hdot F=0,\eqno(4.5)$$
which also has a bilinear form (3.11).

In the case ${\bf N=7}$ the only equation satisfying the Painlev\'e
requirement is the $x$-derivative of the Sawada-Kotera-Ramani
equation,
$$
(3T_x^5T_x^{*2}+5\alpha(T_x^4T_y^*-T_x^3T_x^*T_y)
-15\alpha^2T_xT_y^2-3T_x^2T_z)F\hdot F\hdot F=0,
\eqno(4.6)$$
which means that the original equation is probably
integrable even with an inhomogeneous $k(y,t)$ term.

For ${\bf N=8}$ only the case 8a admits additive terms.  Moreover it
turns out that the first subleading term must also be 1-dimensional
and it coincides with 6a. The full equation in this case reads:
$$\left(4T_x^7T_x^*+5T_x^4T_x^{*4}+\alpha(20T_x^3T_x^{*3}+7T_x^6)+9\beta
T_x^2T_x^*T_y^*+{9\alpha\beta\over 2}T_xT_y^*\right)F\hdot F\hdot
F=0,\eqno(4.7)$$ and its nonlinear form with $u=6\partial_x\log F$ is:
$$\displaylines{u_{7x}+6u_{5x}u_x+10u_{4x}u_{xx}+5u_{xxx}^2+10u_{xxx}u_x^2+{
5\over 3}u_x^4
\hfill\cr +\alpha(3u_{5x}+10u_{xxx}u_{x}+5u_{xx}^2+{10\over 3}u_x^3)
\cr\hfill +\beta(u_{xxy}+u_xu_y)+{\alpha\beta\over 2}u_y=0 \quad(4.8)\cr}$$
To our knowledge this equation is a new integrable PDE.

\noindent Case ${\bf N=10}$:
\noindent Again one equation satisfies the integrability criterion. It
reads:
$$\displaylines{\left(5T_x^8T_x^{*2}+4T_x^5T_x^{*5}+
\alpha(4T_x^7T_x^*+5T_x^4T_x^ {*4})+\beta
(20T_x^3T_x^{*2}T_y^*+7T_x^5T_y\right)\hfill\cr\hfill +6\alpha\beta
T_x^2T_x^*T_y^* +{3\beta^2\over 2}T_yT_y^*)F\hdot F\hdot
F=0\quad(4.9)\cr}$$ This equation, too, has not been encountered
before as an integrable equation.

\vskip 0.6truecm
\noindent{\bf 5. Conclusion}
\medskip
\noindent
In the preceding paragraphs we have presented an extension of Hirota's
bilinear formalism that can encompass any degree of
multilinearity. The main guide in our investigation has been the
requirement that the equations be gauge-invariant. Since our
objective is the study of integrability we have also presented a
classification of one-component trilinear equations that pass the
Painlev\'e test.

An interesting difference between the bilinear and the tri- (and
multi-)linear cases is that now free parameters enter already at the
leading part. In practice this means that the Painlev\'e analysis of
the higher order unicomponent equations becomes increasingly
difficult.  Once the leading parts of these equations are fixed, we
have studied the lower-order terms that can be added without
destroying the Painlev\'e property.  As a result new integrable
equations were discovered. The fact that they have eluded discovery
till now is understandable since these equations are of high order and
there is no hope to discover them by chance: a solid guide and a
systematic approach are needed.

Starting from a complete classification of unicomponent equations one
can build up multicomponent ones following the approach we presented in
[10,11] for the bilinear case.  Further extensions can be presented and we
can, of course, construct also higher multilinear equations (quadri-,
penta-, etc.) equations.

Extension of our formalism to the discrete case is also possible. In
this case the integrability requirement is just the property of
singularity confinement [15]. Some preliminary results exist already
in this direction.

Thus the multilinear extension to Hirota bilinear approach has a wide
range of applicability. There are still many open problems, e.g., the
computation of the multisoliton solutions of the trilinear
equations. Hopefully also the $\tau$-function formalism of the Kyoto
school [16] can be extended in this direction.

\vskip 0.6truecm
\noindent{\bf Acknowledgements}

\noindent  The authors are grateful to R. Hirota, J. Satsuma, W. Oevel
and R. Conte for illuminating discussions and exchange of
correspondence, to W. Strampp for communicating his latest works on
the subject and to J. Springael for comments on the manuscript.

\vfill\eject
\noindent{\bf References}

\medskip
\item{[1]} F. Calogero, in {\sl Applications of analytical and geometric
methods to nonlinear differential equations}, P.A. Clarkson ed., NATO
ASI C413, Kluwer (1993), p.109.
\item{[2]} R. Hirota, Phys. Rev. Lett. {\bf 27} (1971) 1192.
\item{[3]} R. Hirota in {\it Soliton}, R. K. Bullough and P. J. Caudrey
eds., Springer (1980), p. 157.
\item{[4]} M.J. Ablowitz, A. Ramani and H. Segur, Lett. Nuovo Cim. {\bf 23}
(1978) 333.
\item{[5]} B. Grammaticos,  A. Ramani and J. Hietarinta , Phys. Lett.
{\bf A190} (1994) 65.
\item{[6]} J. Matsukidaira, J. Satsuma and W. Strampp, Phys. Lett.
{\bf A147} (1990) 467. A extension of this trilinear approach, that
preserves the $3\times 3$ determinantal form, has been recently
proposed by Strampp and coworkers: Y. Cheng, W. Strampp and B. Zheng:
``Constraints of the KP hierarchy and multilinear forms'',
Comm. Math. Phys., to appear.
\item{[7]} K. Sawada and T. Kotera, Progr. Theor. Phys. {\bf 51} (1974)
1355; A. Ramani, in {\it Fourth International Conference on Collective
 phenomena}, J.L. Lebowitz ed., New York Academy of Sciences (1981)
 p. 54.
\item{[8]} H.C. Morris, J. Math. Phys. {\bf 18} (1977) 530.
\item{[9]} D.J. Kaup, Studies Appl. Math. {\bf 62} (1980) 189;
 B.A. Kupershmidt and G. Wilson, Invent. Math. {\bf 62} (1981) 403.
\item{[10]} J. Hietarinta, J. Math. Phys. {\bf 28} (1987) 1732, 2094,
2586; ibid {\bf 29} (1988) 628.
\item{[11]} B. Grammaticos,  A. Ramani and J. Hietarinta , J. Math. Phys.
{\bf 31} (1990) 2572.
\item{[12]} J. Chazy, Acta Mathematica, {\bf 34} (1911) 317.
\item{[13]} A. Hearn, REDUCE User's manual, Version 3.4 (Rand, Santa Monica,
1991).
\item{[14]} J. Schiff, in {\it Painlev\'e Transcendents, Their Asymptotics
and Physical Applications}, D. Levi and P. Winternitz eds., Plenum
(1992), p. 393.
\item{[15]} A. Ramani, B. Grammaticos and J. Satsuma, Phys. Lett. {\bf A169}
(1992) 323.
\item{[16]} M. Jimbo and T. Miwa, Publ. Res. Inst. Math. Sci {\bf 190} (1983)
943.

\end